\documentclass{elsart}

\usepackage{psfig}

\usepackage{natbib}

\begin{document}

\runauthor{Kiyoshi Hayashida}


\begin{frontmatter}

\title{ASCA Observations of NLS1s: BH Mass Estimation from X-ray Variability
and X-ray Spectra}

\author[OSAKA]{K. Hayashida}
\address[OSAKA]{Department of Earth \& Space Science, Osaka
University,
1-1 Machikeneyama, Toyonaka, Osaka 560-0043, Japan}

\begin{abstract}
ASCA observations of Narrow-Line Seyfert 1 galaxies (NLS1s) are presented.
We focus on the black hole size of the NLS1 sources by employing
two independent methods for the mass estimation; one is using X-ray
variability, the other is using a blackbody fit to the soft component.
Although the coincidence is not good for some sources, the mass
estimated by these methods ranges from $10^5$ to $10^7$ $M_\odot$, 
systematically smaller than those for typical (broad line) Seyfert 1.
We consider the small mass black hole to be the principal cause of
the several extreme characteristics of the NLS1s.
\end{abstract}


\end{frontmatter}


\section{Introduction}
Properties of Narrow Line Seyfert 1 type (NLS1)
sources are summarized as, 1) steep soft X-ray 
spectrum, 2) rapid X-ray variability, 3) narrow H$\beta$ width
(see {\it e.g.} Boller {\it et al.} 1996). 
These properties and their distinctive differences from Broad 
Line Seyfert 1 type (BLS1) galaxies are expected to be explained by 
fundamental parameters of AGNs, such as geometry, black hole (BH)
mass, accretion rate, and so on. 
In this paper, we try to estimate the BH mass of the NLS1s (and 
BLS1s) by two independent methods based on the ASCA observations;
One is from X-ray variability of the sources, and the other is 
from a black body fit to the soft X-ray spectrum.  
We suggest that the NLS1s have smaller black holes than usual
Seyfert 1s and emit X-rays  with a higher efficiency. 

Earlier results with a smaller sample was presented in 
Hayashida (1998, 2000). In this paper, 
we treat 14 NLS1 sources, 
IZw1, Ton S 180, PHL 1092, PKS 0558-504, 1H 0707-495,
RE 1034+39, NGC 4051, PG 1211+143, Mrk 766,
PG 1244+226, IRAS 13224-3809, PG 1404+226, Mrk 478 and Ark 564. 
We also use 9 BLS1 sources observed with Ginga and ASCA
for the variability analysis.  Among the nine BLS1s, we include  
MCG-6-30-15. However, the source may better be classified 
as a NLS1 or a narrow emission line galaxy.  We adopt $H_0=75km/s/Mpc$
in this paper.

\section{Black Hole Size in the NLS1s}
\subsection{X-ray Variability}
Rapid and large amplitude X-ray variability is one of the common 
features of NLS1s.  The most extreme case was found in IRAS13224 
(Otani {\it et al.} (1996) and Boller {\it et al.} (1997)); the maximum
variability was a factor of 60, and the shortest doubling time was 800~s. 

\begin{figure}[b]
\centerline{
\psfig{figure=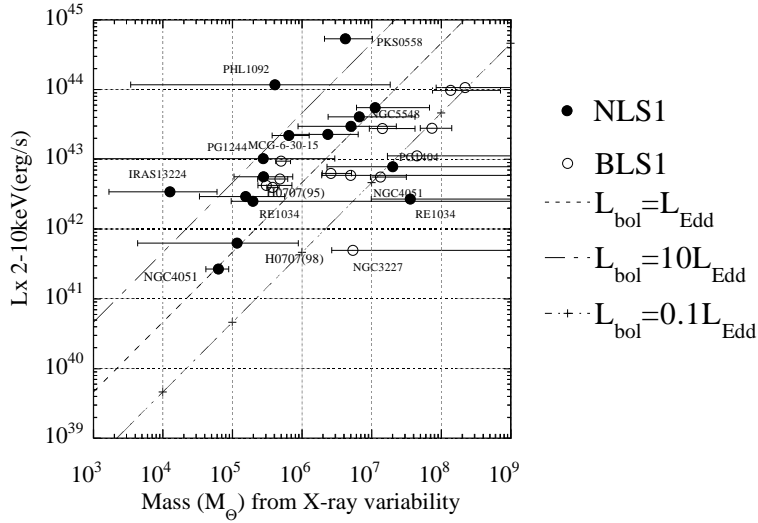,height=7cm,angle=0}}
\caption{X-ray(2-10keV) luminosity versus black hole mass estimated
from the X-ray variability. Closed circles represent NLS1s and 
open circles, BLS1s. The lines for the Eddington ratio are calculated 
assuming a bolometric correction of 27.2,  which is appropriate
for typical Seyferts, but may not be appropriate for NLS1s.}
\end{figure}

We have investigated the X-ray variability of Seyfert 1 galaxies,
in which we developed a method to extract relative variability 
time scale of sources using a normalized power spectrum (Hayashida {et 
al.} 1998a). 
Another point of our work was that, employing the stellar black 
hole Cyg X-1 as the base point, and assuming the variability scales 
linearly with the time scale and the system size, we estimate
the central black hole of AGNs. We applied the same procedure
to the NLS1s and the BLS1s in our sample (Fig.1).
As shown in Fig.1, our NLS1 sample have black hole masses ranging 
from  $10^5$ $M_{\odot}$ to $10^7$ $M_{\odot}$, systematically 
lower than the estimated masses of BLS1s plotted in the same figure. 
In terms of the Eddington ratio, the NLS1s have typical values of 
1-10, while the BLS1s have 0.1-10.  Differences in the estimated
BH mass in NLS1s and in BLS1s are apparent if we look at the correlation with
FWHM of $H_\beta$ (Fig.2).

\begin{figure}[t]
\centerline{
\psfig{figure=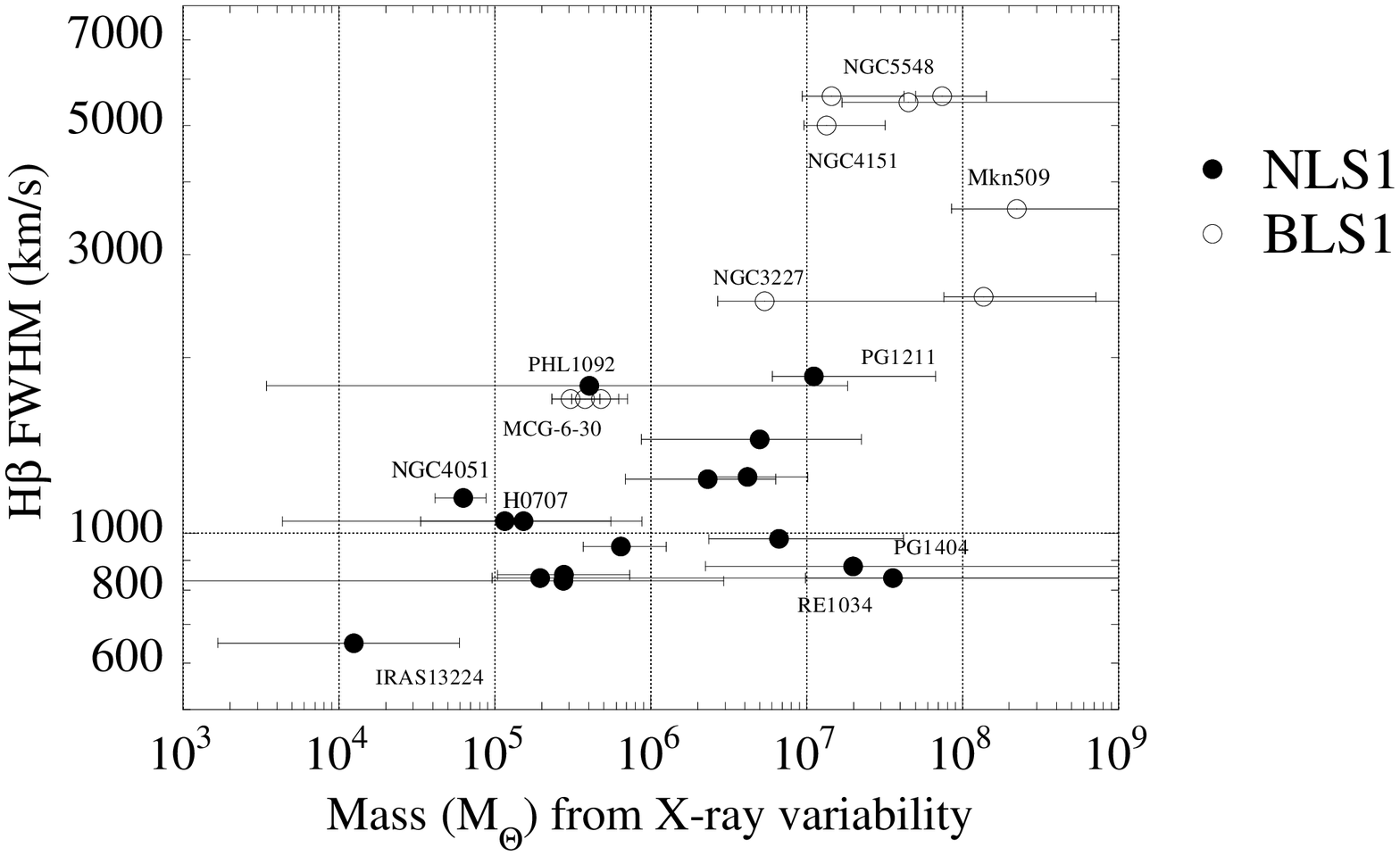,height=5cm,angle=0}
}
\caption{FWHM of $H_\beta$ vs BH mass estimated from the X-ray variability.
}
\end{figure}

\subsection{Blackbody Model Fit to the Soft Component}
One of the most plausible interpretations of the soft component 
of NLS1s is thermal emission from the accretion disk. 
When we fit the soft component with a blackbody model, we obtain
the area of the emission region.  
If we assume the emission region to be the surface of 
a 3 Schwarzshild radius ($R_S$) sphere or a 0.5 $R_S$ sphere,
the black hole size can be estimated. 
Fig.3 shows the luminosity of the blackbody component versus
the mass of the black hole estimated in this way.  If we assume
the emission area size ($r_{bb}$)=3$R_{S}$, most of the 
sources exceed the Eddington limit. On the other hand, if 
$r_{bb}$=0.5$R_{S}$, i.e. the extreme Kerr case, the radiation is at the 
sub-Eddignton phase. A more important thing is that the  mass ranges 
from $10^5$ to $10^6$ $M_\odot$. 

\begin{figure}[b]
\centerline{
\psfig{figure=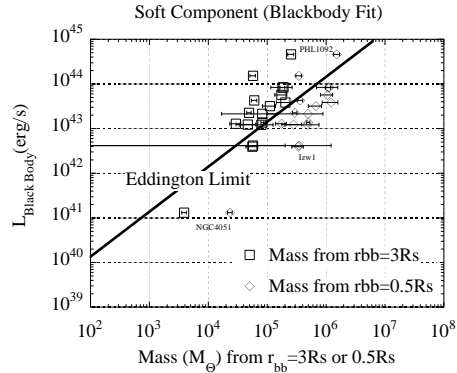,height=5cm,angle=0}
}
\caption{Luminosity of the blackbody component versus Mass
estimated from the spectral fit for two different assumptions, {\it i.e.}, 
$r_{bb}$=3$R_{S}$ or 0.5$R_{S}$. 
}
\end{figure}

\subsection{Comparison of Two Mass Estimates}
Both mass estimates are compared directly in 
Fig. 4.  A large discrepancy of up to two orders of magnitude
is found for some sources. However, for the sources below the 
diagonal line, we have a possible reason for it; 1) we neglected 
the disk inclination factor, with which the data points should move upward,
and 2) the color temperature we used might be higher than the effective
temperature that we should have used, if we had a sophisticated model
for the disk emission. On the other hand, these facts cannot
reconcile the discrepancy for the sources above the diagonal line in 
Fig.4. Nevertheless, we find that most of those sources have an 
extremely enhanced soft component relative to the hard power
law component. One of the speculations we have is that those 
sources are in some extreme state in which the >variability mode 
is very much different from the others, {e.g.}, intrinsic amplitude 
is enhanced. 

\begin{figure}[h]
\centerline{
\psfig{figure=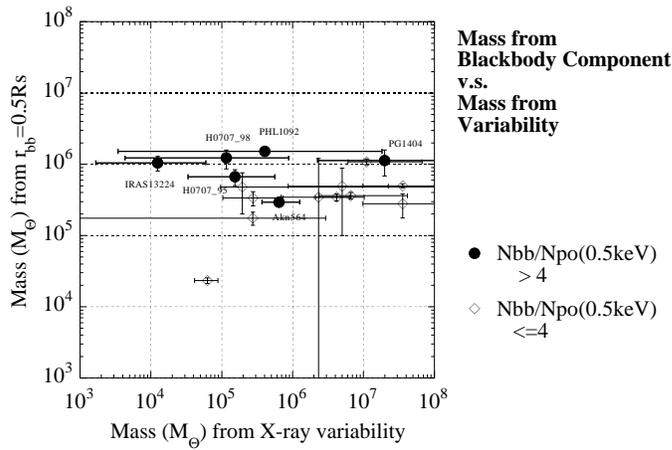,height=6cm,angle=0}
}
\caption{Comparison between the mass from X-ray variability and the 
mass from the blackbody fit with the 0.5$R_{S}$ assumption.
Note that if we adopt 3$R_{S}$,  the data points move downward
by a factor of 6. We distinguish NLS1s by the intensity 
ratio of their soft black body and hard power law spectral components 
at 0.5 keV. }
\end{figure}





\end{document}